\documentclass[12pt]{article}
\usepackage{graphicx}

\newcommand{\be}{\begin{equation}}
\newcommand{\dd}{\displaystyle}
\newcommand{\ee}{\end{equation}}
\newcommand{\bea}{\begin{eqnarray}}
\newcommand{\eea}{\end{eqnarray}}

\newcommand{\nn}{\nonumber}
\newcommand{\de}{\partial}
\begin{document}
\hfill{\bf BARI-TH 438/02}\par
 \hfill{\bf UGVA-DPT-2002 05/1100}

\hfill$\vcenter{
  \hbox{\bf}
 \hbox{\bf } }$
 \begin{center}
 {\Large\bf\boldmath { Crystalline Color Superconductivity: Effective Lagrangian and Phonon Dispersion Law}}
 \rm \vskip1pc {\large
 R. Casalbuoni$^{a,b}$, R. Gatto$^c$
 and G. Nardulli$^{d,e }$}\\ \vspace{5mm} {\it{$^a$Dipartimento di
 Fisica, Universit\`a di Firenze, I-50125 Firenze, Italia
 \\
 $^b$I.N.F.N., Sezione di Firenze, I-50125 Firenze, Italia\\
 $^c$D\'epart. de Physique Th\'eorique, Universit\'e de Gen\`eve,
 CH-1211 Gen\`eve 4, Suisse\\ $\dd ^d$Dipartimento di Fisica,
 Universit\`a di Bari, I-70124 Bari, Italia  \\$^e$I.N.F.N.,
 Sezione di Bari, I-70124 Bari, Italia }}
 \end{center}
 \begin{abstract}
The recent study by Bowers and Rajagopal on the possible
structures for the QCD crystalline color superconducting phase
favors a  cubic structure and allows for a first discussion of the
QCD dynamics in the crystalline phase. The cubic structure leads
to complete breaking of traslation invariance and to three phonon
modes. We discuss the expected form of the effective Lagrangian
and the phonon dispersion laws assuming the proposed crystal
structure. Besides the interest for QCD and the related possible
relevance to compact stars these considerations may become useful
to atomic systems exhibiting crystalline superfluidity.
 \end{abstract}
\section{Introduction}

In a recent paper Bowers and Rajagopal \cite{bowers} have presented
a comprehensive study of the different possible crystal structures
for the crystalline color superconducting phase of QCD and
concluded in favor of a face centered cubic condensate pattern.
This study now allows discussion of the dynamics of QCD
in the crystalline phase and this note is a first contribution
in that direction.

The subject of the color superconducting phases of QCD
\cite{barrois} \cite{alford} has only recently had its most
extensive developments \cite{review}. At asymptotically large
values of the chemical potential for quark number $\mu$, cold QCD
with three degenerate flavors lives in the color-flavor locked
(CFL) phase \cite{arw} corresponding to condensation of quark
pairs of opposite momenta and lying near the Fermi surfaces (BCS
pairing).

For finite strange quark mass the CFL phase
is expected to persist within some range below asymptotic densities.
Decreasing the value of $\mu$, the expected possibility becomes
pairing between quarks of non vanishing total momentum, each quark
momentum lying near the respective (now differing) Fermi surfaces
\cite{abr}  \cite{bkrs} \cite{casa} \cite{leib} \cite{reviewsloff}. Closeness
to the
Fermi surface will count for a lower free energy. Quantitative studies so far
are mostly for two flavors and a difference in their chemical potential
$\delta\mu$ is usually introduced to induce the difference in the Fermi
momenta.

Similar possibilities of fermion pairing of non vanishing total
momentum were already encountered in condensed matter
(ferromagnetic systems with paramagnetic impurities) \cite{loff},
so that the name LOFF phase is often used, from the initials of
the authors of ref. \cite{loff}.

In a LOFF phase translational and rotational invariance are
spontaneously broken. The most relevant question for crystalline
superconductivity is that of the space dependence of the Cooper
pair condensate. Preliminary studies have been based on a plane
wave ansatz: a single plane wave with the finite momentum of the
Cooper pairs, assuming that all the momenta spontaneously choose a
unique direction. These studies (see ref \cite{abr} and
\cite{loff}) have lead to a picture of the LOFF phase existing
within a certain window of $\delta\mu$ which starts with a first
order phase transition and ends at the upper end with a second
order transition.

The authors of \cite{bowers} develop a Ginzburg-Landau expansion
near the second order transition. They introduce a multiple
plane-wave ansatz and find that the favorite crystal structure is
that of a face-centered cubic crystal.

For the simple single plane wave structure, so far adopted as a
provisional working hypothesis, we had constructed \cite{casa} the
effective low energy phonon lagrangian and derived the phonon
dispersion law. In that case only one phonon was expected from the
partial breaking of translation invariance. For the cubic
structure now derived in \cite{bowers} one will have three
phonons, due to the complete breaking of translation invariance.

The work \cite{bowers} uses a simplified model with pointlike
interaction and two massles quarks and, as we have said, a
Ginzburg-Landau approximation. The lowest value of the
Ginzburg-Landau free energy corresponds to a crystal structure
built up from eight plane waves directed to the corners of a cube.
In spite of the approximations used one has the qualitative
feeling that their proposed crystal structure would remain the
favorite one also in a more general QCD setting (although not
perhaps at asymptotic densities).

In this note we shall discuss the expected form of the effective
Lagrangian and phonon dispersion laws assuming the cubic crystal
structure of \cite{bowers}.

We have already mentioned the theoretical interest of color
supeconductivity in the crystalline phase as one of the general
components of the QCD phase diagram. Physically, crystalline
color superconductivity might be realized within compact stellar
systems. In fact the quark chemical potential is expected to
increase in penetrating towards the center of a compact star and
other parameters, such as the strange quark mass, are also
expected to evolve in such a way that one may reach the
crystalline phase inside a core region of the star. In a rotating
compact star the ensuing vortices may give rise to observable
glitches, as discussed in ref. \cite{abr}.

The considerations developed in connection to crystalline
superconductivity in QCD, including those presented in this note,
may also be useful in the field of atomic systems, in particular,
as mentioned in ref. \cite{bowers}, for gases of fermionic atoms
below their Fermi temperature \cite{screck} \cite{holland}. These
systems present over QCD the advantage of a better parametric
flexibility \cite{combescot}.

\section{The LOFF phase of QCD}

According to \cite{bowers} the favored crystalline structure in
the LOFF phase of QCD should be a face-centered-cubic one. This is
characterized by eight $\vec q_j$ vectors  with equal length
$q=\pi/a$ and directions corresponding to the eight vertices of
the cube in Fig. 1: \be \Delta(\vec
x)=\Delta\sum_{j=1}^8\,\exp\{2i\vec q_j\cdot\vec
x\}=\Delta\sum_{\epsilon_i=\pm}\exp \left\{\dd{\frac{2\pi
i}a}(\epsilon_1 x_1+\epsilon_2 x_2+\epsilon_3 x_3)\right\}\ .
\label{1}\ee

\begin{center}
\includegraphics*[scale=.35]{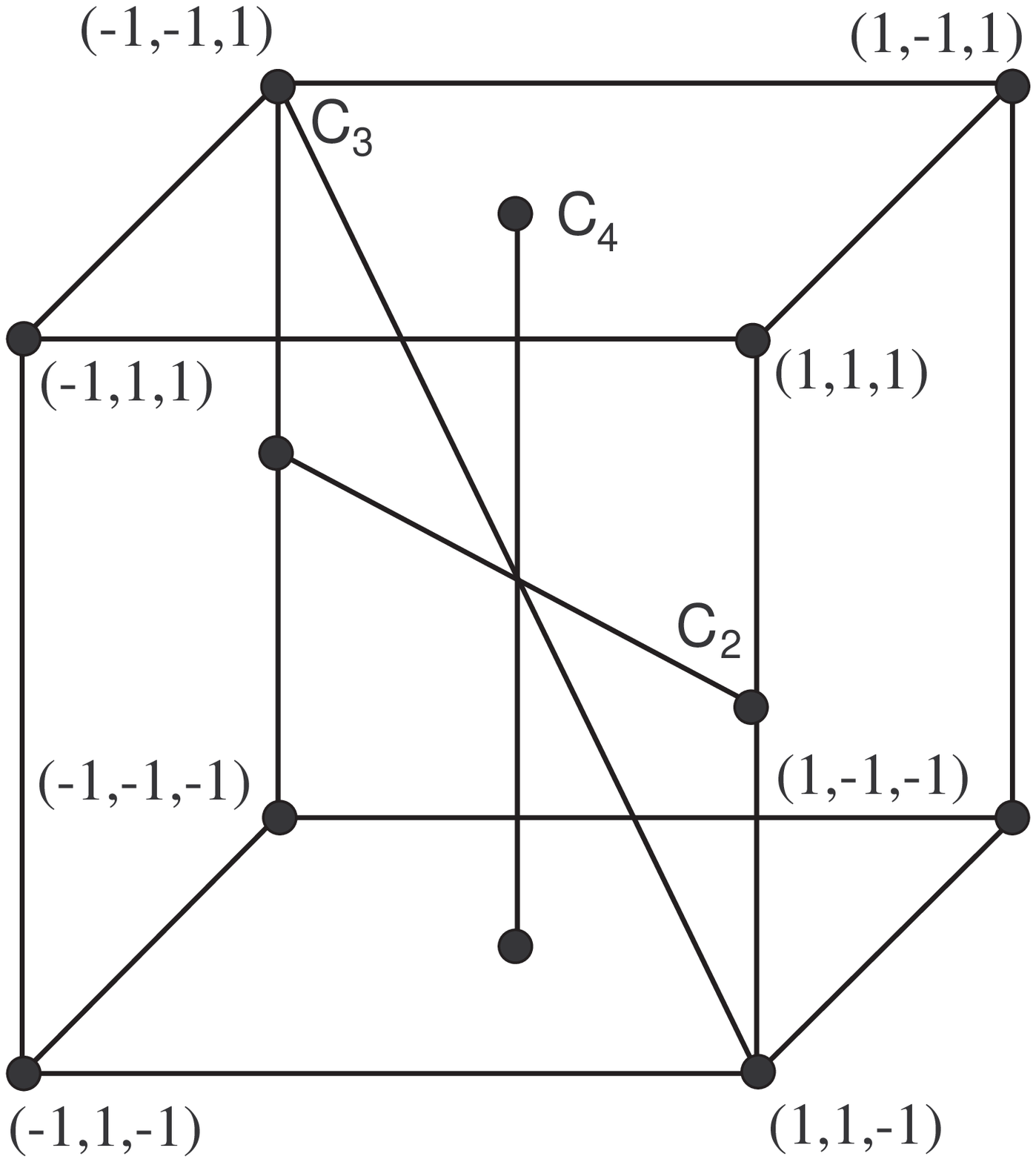}
\end{center}\noindent Fig. 1 - {\it{The figure shows  some of the
symmetry axes of the cube, denoted as  $C_2$, $C_3$ and $C_4$ (see
the text). We show also the parameterization used for the
coordinates of the vertices.}}\\

In Fig. 1 some of the symmetry axes of the cube are shown: they
are denoted as $C_4$ (the three 4-fold axes), $C_3$ (the four
3-fold axes) and $C_2$ (the six 2-fold axes). For more details on
the symmetry group of the cube see  the book by Hamermesh
\cite{hamermesh}.

We observe that the coordinates at the vertices are nothing but
the directions of the $\vec q_j$ vectors. The condensate (\ref{1})
can be also written as follows:
 \bea &\Delta(\vec
x)=2\Delta\Big[\cos\dd{\frac{2\pi}a}(x_1+x_2+x_3)+
\cos\dd{\frac{2\pi}a}(x_1-x_2+x_3)\nn\\&+
\cos\dd{\frac{2\pi}a}(x_1+x_2-x_3)+
\cos\dd{\frac{2\pi}a}(-x_1+x_2+x_3)\Big]\ . \eea This condensate
breaks both translations and rotations. It is however invariant
under the discrete group $O_h$, the symmetry group of the cube.
This can be seen  by noticing that the previous expression is
invariant under the following coordinate transformations \bea
&&R_1:~~~ x_1\to x_1,~~~x_2\to x_3,~~ x_3\to -x_2,\nn\\
&&R_2:~~~ x_1\to -x_3,~~~x_2\to x_2,~~~ x_3\to x_1,\nn\\
&&R_3:~~~x_1\to x_2,~~~x_2\to -x_1,~~~x_3\to x_3,\nn\\ &&I:~~~~~
x_1\to -x_1,~~~x_2\to -x_2,~~~x_3\to -x_3, \label{basic}\eea that
is rotations of $\pi/2$ around the coordinate axes, and the
inversion with respect to the origin. Since one can easily check
that the group $O_h$ is generated by the previous 4 elements  the
invariance follows at once.

We can then introduce 3 scalar fields $\Phi^{(i)}$ such that \be
\langle\Phi^{(i)}\rangle_0=\frac{2\pi}a x_i\label{vacuum}\ee These
fields should be defined on the lattice sites, but for the
purpose of writing the effective action for such fields, we are
interested only in the low energy behavior $p\ll 1/a$. Therefore
we will treat them as continuous fields. The coupling of the
Fermi fields generated by the condensate will be written as
 \be
 \Delta\psi^TC\psi\sum_{\epsilon_i=\pm}\exp\left\{
i(\epsilon_1\Phi^{(1)}+\epsilon_2\Phi^{(2)}+\epsilon_3\Phi^{(3)})\right\}
 \label{coupling}\ee
making the theory invariant under translations and rotations.
These invariances are broken spontaneously in the vacuum defined
by eq. (\ref{vacuum}).

The effective lagrangian for the fields $\Phi^{(i)}$ has to enjoy
the following symmetries: rotational and translational invariance;
 $O_h$ symmetry on the fields $\Phi^{(i)}$.
The latter requirement follows from the invariance of the coupling
(\ref{coupling}) under the group $O_h$ acting upon $\Phi^{(i)}$.

We then define \be \Phi^{(i)}(x)=\frac{2\pi} a
x^i+\phi^{(i)}(x)/f\,,\label{small}\ee with \be
\langle\phi^{(i)}(x)\rangle_0=0\ .\ee and $f$ the decay constant
of the phonon with dimension of an energy. It follows that, after
the breaking, the phonon fields $\phi^{(i)}(x)$ and the
coordinates $x^i$ must transform under the diagonal discrete
group obtained from the direct product of the rotation group
acting over the coordinates and the $O_h$ group acting over
$\Phi^{(i)}(x)$. This is indeed the symmetry left after the
breaking of translational and rotational invariance.

In order to build up the effective lagrangian, we start noticing
that if $X_i$ are quantities transforming as the same
representation of the $\Phi^{(i)}$'s under $O_h$ (a 3-dimensional
representation), then all the invariant expressions can be
obtained by the following three basic invariant expressions \be
I_2(X_i)=X_1^2+X_2^2+X_3^2,~~I_4(X_i)=X_1^2 X_2^2 +X_2^2 X_3^2+
X_3^2 X_1^2,~~ I_6(X_i)=X_1^2 X_2^2 X_3^2.\ee The invariance is
easily checked by noticing that these expressions are invariant
under the three elementary rotations $R_i$  and the inversion $I$
(see eq. (\ref{basic})). Quantities transforming as $\Phi^{(i)}$
are \be \dot\Phi^{(i)}\equiv \frac{d\Phi^{(i)}}{dt}.~~~~
\vec\nabla\Phi^{(i)}\ .\ee We proceed now as in ref. \cite{casa},
by noticing that the most general low-energy effective lagrangian,
invariant under rotations, translations and $O_h$ is given by \be
{\cal L}=\frac {f^2} 2\sum_{i=1,2,3}({\dot\Phi}^{(i)})^2+ {\cal
L}_{\rm s}(I_2(\vec\nabla\Phi^{(i)}),
I_4(\vec\nabla\Phi^{(i)}),I_6(\vec\nabla\Phi^{(i)}))\ .\ee The
reason for allowing all the possible spatial gradients in this
expansion is that the vacuum expectation value of
$\vec\nabla\Phi^{(i)}$ is given by $x_i$ and it is not in general
small.

Define the auxiliary quantities \be
X_i(\epsilon)=\frac{2\pi}a{\vec
e\,}^{(i)}+\epsilon\vec\nabla\phi^{(i)}\ee such that \be
X_i(1/f)=\vec\nabla\Phi^{(i)}\ ,\ee where ${\vec e\,}^{(i)}$ is a
unit vector along the axis $"i"$. The $\epsilon$ parameter has
been introduced to keep track of the spatial derivatives. Since we
are interested at the second order in these derivatives we need
the quantity (first order terms are total derivatives)\be \frac 1
2 \frac{\de^2{\cal L}(\epsilon)}{\de\epsilon^2}\Big|_{\epsilon=0}=
\frac 12\left[\sum_{i,j=2,4,6}\frac{\de^2{\cal L}}{\de I_i\de
I_j}\frac{\de I_i}{\de\epsilon}\frac{\de I_j}{\de\epsilon}+
\sum_{i=2,4,6}\frac{\de{\cal L}}{\de I_i}\frac{\de^2
I_i}{\de\epsilon^2}\right]_{\epsilon=0}\label{97}\ee where ${\cal
L}(\epsilon)$ has been obtained by the substitution
$\vec\nabla\Phi^{(i)}\to  X_i(\epsilon)$. Since the derivatives of
${\cal L}$ with respect to the invariant terms $I_i$, evaluated at
$\epsilon=0$ are just constant, using \be \frac{\de
I_2}{\de\epsilon}\Big|_{\epsilon=0}=
2\sum_{i=1,2,3}\de_i\phi^{(i)},~~\frac{\de
I_4}{\de\epsilon}\Big|_{\epsilon=0}=
4\sum_{i=1,2,3}\de_i\phi^{(i)},~~\frac{\de
I_6}{\de\epsilon}\Big|_{\epsilon=0}=
2\sum_{i=1,2,3}\de_i\phi^{(i)}\ee and \bea \frac 1 2 \frac{\de^2
I_2}{\de
\epsilon^2}\Big|_{\epsilon=0}&=&\sum_{i=1,2,3}|\vec\nabla\phi^{(i)}|^2\nn\\
\frac 1 2 \frac{\de^2 I_4}{\de
\epsilon^2}\Big|_{\epsilon=0}&=&4\sum_{i<j=1,2,3}\de_i\phi^{(i)}\de_j\phi^{(j)}+
2\sum_{i=1,2,3}|\vec\nabla\phi^{(i)}|^2 \nn\\
\frac 1 2 \frac{\de^2 I_6}{\de
\epsilon^2}\Big|_{\epsilon=0}&=&4\sum_{i<j=1,2,3}\de_i\phi^{(i)}\de_j\phi^{(j)}+
\sum_{i=1,2,3}|\vec\nabla\phi^{(i)}|^2\,, \eea we see that the
expression (\ref{97}) depends only on the quantities \be
\sum_{i=1,2,3}|\vec\nabla\phi^{(i)}|^2,~~~
\sum_{i=1,2,3}(\de_i\phi^{(i)})^2,~~~
\sum_{i<j=1,2,3}\de_i\phi^{(i)}\de_j\phi^{(j)}\,.\ee

In conclusion one gets \be {\cal L}=\frac 1
2\sum_{i=1,2,3}({\dot\phi}^{(i)})^2-\frac a 2
\sum_{i=1,2,3}|\vec\nabla\phi^{(i)}|^2- \frac b 2
\sum_{i=1,2,3}(\de_i\phi^{(i)})^2-
c\sum_{i<j=1,2,3}\de_i\phi^{(i)}\de_j\phi^{(j)}\,.\ee This
expression could have been guessed by noticing that after the
breaking one still has an invariance under the group $O_h$ acting
at the same time upon the coordinates and the phonon fields
$\phi^{(i)}$.
\section{Dispersion laws}

To analyze the phonon dispersion law one has to diagonalize the
matrix \be M= \left( \matrix{ a\,|\vec p\,|^2 +b\,p_1^2 & c\,p_1
p_2 & c\, p_1 p_3 \cr c\, p_1 p_2 & a\,|\vec p\,|^2 + b\,p_2^2 &
c\, p_2 p_3\cr c\, p_1 p_3 & c\, p_2 p_3 & a\,|\vec p\,|^2
+b\,p_3^2 } \right)\label{matrix}\,. \ee

The dispersion law can be written as follows ($\vec p=\vec n\,
p$): \be E=v_r(\vec n)p\ \ee where $v_r^2(\vec n) $ are given by
\be v_r^2(\vec n)=a+\frac b 3 -u_r+\frac{\dd{J_4-\left(\frac b 3
\right)^2}}{u_r}\ee and $u_r$ are the three cubic roots defined by
\be u_r(\vec n)=\sqrt[3]{A(\vec n)+\sqrt{B(\vec n)}} \ .\ee Here
\bea A(\vec n)&=&\frac{bJ_4-J_6}2\,-\, \left(\frac b 3\right)^3\cr
B(\vec n)&=&J_4^3-\frac{b^2}{12}J_4^2-\frac{b}2\,J_4J_6\,+\,\left(
\frac{b}{3}\right)^3\,J_6+\frac{J_6^2}{4}\eea and \bea J_4 &=&
\frac{b^2-c^2}{3}I_4(\vec n)\cr
 J_6 &=& (b+2c) (b-c)^2 I_6(\vec n)\ .\eea
These solutions display the same symmetries of the effective
lagrangian   because they depend on the directions $\vec n$ only
through the elementary symmetric functions $I_2(\vec
n)=1,\,I_4(\vec n),\, I_6(\vec n)$. The corresponding
eigenvectors can be computed in a straightforward manner; for
example if \be n_1\neq 0\ ,\hskip1cm a-v_r^2+bn_3^2\neq 0\
,\hskip1cm a-v_r^2+(b-c)n_2^2\neq 0\label{conditions}\ee one gets
the eigenvectors: \be \xi_r\ =\ {\rm const.}\times
 \left(\matrix{
   n_1  [a- v^2_r+n_2^2(b-c) ]\cr
   \cr
   n_2 [a-v^2_r+n_1^2(b-c)] \cr\cr
   -n_3 c\,{\dd \frac{(a-v^2_r)(n_1^2+n_2^2)+2(b-c)n_1^2n_2^2}
   {a-v^2_r+b\,n_3^2}}}\right)\label{19}\ .
 \ee
The dispersion law for the phonons depends on three arbitrary
parameters $a,\,b,\,c$. This follows from the symmetry of the
cubic crystal and
 could be also inferred from the general theory of the elastic waves
 in crystals. As matter of fact, the
 phonon amplitudes $\phi^{j}$  can be identified with the local
 deformations of the crystal with harmonic components
 \be\varphi_j=\varphi_{0\,j}\,e^{i(\vec p\cdot\vec x-Et)}\ .\ee
 The equation of motion for the deformation is
\be \rho\frac{\partial^2 \varphi_j}{\partial t^2}=
\partial_k\sigma_{jk}\ee
where $\sigma_{j\,k}$ is the stress tensor and $\rho$ is the
crystal density. The tensor $\sigma_{j\,k}$ is related to the strain tensor
$u_{\ell m}$: \be u_{\ell m}\,=\,\frac 1 2 \left(
\partial_\ell\varphi_m+
\partial_m\varphi_\ell+
\partial_\ell\varphi_i
\partial_m\varphi_i\right) \ ,\ee
by the relation \be \sigma_{i\,k}\,=\,\lambda_{ik\ell m}u_{\ell
m}\ .\ee $\lambda_{ik\ell m}$ is the elastic modulus tensor and
the equation for elastic waves can be written as \be
\rho\frac{\partial^2 \varphi_j}{\partial t^2}= \lambda_{jk\ell m}
\partial_k\partial_\ell\varphi_m\ ,
\ee from which the dispersion law follows: \be Det\left(\rho
E^2\delta_{im}- \lambda_{ik\ell m}u_{\ell m}p_kp_\ell \right)=0\
.\ee Owing to its symmetry properties, the elastic modulus tensor
depends in general on 21 parameters; for the cube group $O_h$
only three of them are independent \cite{LL}, for example
$\lambda_{1111},\,\lambda_{1122},\,\lambda_{1212},$ The
identification with the parameters of the effective lagrangian is
as follows: \be a=\frac{\lambda_{1212}}{\rho},~~~
b=\frac{\lambda_{1111}\,-\,\lambda_{1212}}{\rho},~~~
c=\frac{\lambda_{1122}\,+\,\lambda_{1212}}{\rho}. \ee

The group velocity
\be\vec v=\frac{\partial E}{\partial \vec p}\ee
is a function of the direction of the wave propagation $\vec n$ but does
not depend of the energy. For each fixed $\vec n$ there are three
velocities that are in general all different. For comparison in an
isotropic medium $|\vec v|$ does not depend on $\vec n$  and two
of the velocities (the transverse ones) coincide, while being
 different and
smaller than
the  velocity of the longitudinal waves.

Since the matrix $M$ in (\ref{matrix}) is real and symmetric, its
eigenvalues are real as well. One has also to impose that
$v^2_r>0$; for this to happen all the principal minors \bea
\Delta_1&=&a+bn_i^2\ ,\hskip 6cm(i=1,2,3)\label{d1}
\\
\Delta_2&=&(a+bn_i^2)(a+bn_j^2)-c^2n_i^2n_j^2\ , \hskip 3cm(i\neq
j)\label{d2}
\\
\Delta_3&=&a^2(a+b)+a(b^2-c^2)I_4(\vec n)+(b+2c)(b-c)^2I_6(\vec n)
\hskip 1cm\label{d3} \eea must be positive: \be \Delta_i>0\hskip
2cm(i=1,2,3)\label{29} \ .\ee The constraint $\Delta_1>0$ gives
the conditions \be a>0\ ,\hskip2cm a+b>0\ .\ee The most stringent
constraint $\Delta_2>0$ is obtained for $n_i^2=n_j^2=1/2$, which
gives\be b+2a>|c|\ .\ee Finally the most stringent constraint
$\Delta_3>0$ is obtained for $I_4(\vec n)=1/3$ and $I_6(\vec
n)=1/27$, i.e. when $\vec n$ is parallel to one of the 3-fold
axes; since $\Delta_3$ is the product of the three eigenvalues,
one gets \be a+\frac{b+2c}3>0\ .\ee These constraints are
summarized in Fig. 2.

Let us specialize these results for momenta along the symmetry
axes of the crystal i.e. along the three  4-fold axes, the four
3-fold axes and the six 2-fold axes.

We easily get for $\vec n$ along one of the three  4-fold axes,
e.g. $\vec n= (1,0,0)$, \bea
v_1^2&=&a+b\label{20.1}\\
v_2^2&=&v_3^2=a\label{20.2} \eea and the corresponding
eigenvectors are \be\xi_1=\phi^{(1)} \ee while $\xi_2,\,\xi_3$
are in the subspace spanned by $\phi^{(2)}$ and $\phi^{(3)}$. The
eigenvalues for the other two 3-fold axes with $\vec n= (0,1,0)$
and $\vec n= (0,0,1)$ are identical to (\ref{20.1}) and
(\ref{20.2}) due to the cubic symmetry, while the corresponding
eigenvectors are obtained by cyclic permutations.

\begin{center}
\includegraphics*[scale=.5]{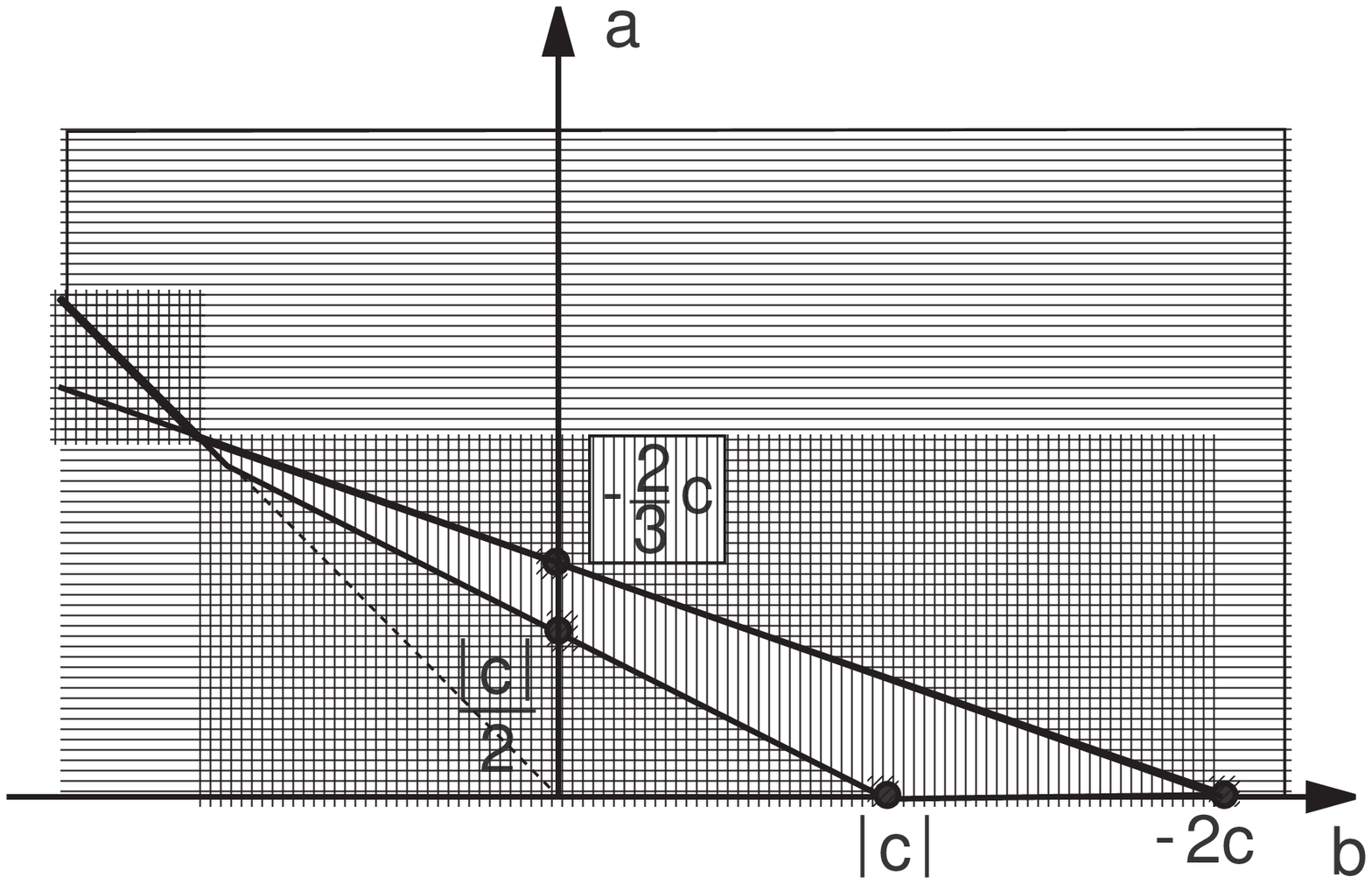}
\end{center}\noindent Fig. 2 - {\it{The figure shows the regions where  the
square of all the velocities of propagation along the symmetry
axes of the crystal are positive, in the plane $(b,a)$ at fixed
$c$. For $c<0$ only the area with horizontal lines is allowed. For
$c>0$ both regions are permitted.}}\\

For $\vec n$ along one of the four 3-fold axes, e.g. $\dd \vec n=
\frac{1}{\sqrt 3}(1,1,1)$ one has \bea v_1^2&=&a+\frac{b+2c}3\cr
v_2^2&=&v_3^2=a+\frac{b-c}3\label{eig2} \eea with eigenvectors
\be\xi_1=\frac{1}{\sqrt 3}\left(\phi^{(1)}+\phi^{(2)}+
\phi^{(3)}\right)\label{eig3} \ee while $\xi_2,\,\xi_3$ are in
the orthogonal subspace,i.e. the subspace spanned by
\be-\phi^{(1)}+x\phi^{(2)}+(1-x)\phi^{(3)}\ ,\label{eig4}\ee with
$x$ a real parameter.

The other three 3-fold  axes are defined by the directions
$\dd \vec n= \frac{1}{\sqrt 3}(1,-1,1)$, $\dd \vec n= \frac{1}{\sqrt
3}(-1,1,1)$
and $\dd \vec n= \frac{1}{\sqrt 3}(-1,-1,1)$; they can be obtained
by the unit vector $\dd \vec n= \frac{1}{\sqrt 3}(1,1,1)$
by applying transformations in $O_h$;
due to the cubic symmetry the eigenvalues are again given by
 (\ref{eig2}) and the corresponding eigenvectors are obtained
 by applying the same transformation  to the eigenvectors
 (\ref{eig3}), (\ref{eig4}); for example for
$\dd \vec n= \frac{1}{\sqrt 3}(1,-1,1)$ one has
\be\xi_1=\frac{1}{\sqrt 3}\left(\phi^{(1)}-\phi^{(2)}+
\phi^{(3)}\right)\label{eig5}
\ee while $\xi_2,\,\xi_3$ are in the orthogonal subspace
 spanned by
\be
\phi^{(1)}+x\phi^{(2)}+(1-x)\phi^{(3)}\ .\label{eig6}\ee

Finally we consider the six 2-fold axes defined by the directions
$\dd \vec n= \frac{1}{\sqrt 2}(1,\pm 1,0)$,
$\dd \vec n= \frac{1}{\sqrt 2}(1,0,\pm 1)$
and $\dd \vec n= \frac{1}{\sqrt 2}(0,0,\pm 1)$.
By way of example we
 take
\be \vec n= \frac{1}{\sqrt 2}(1,\pm 1,0)\label{n}\ .\ee
 The eigenvalues are
\bea
v_1^2&=&a+\frac{b+c}2\label{eig7.0}\\
v_2^2&=&a+\frac{b-c}2\label{eig7.1}\\
v_3^2&=&a\label{eig7.3}\ . \eea The eigenvector corresponding to
the eigenvalue (\ref{eig7.0}) can be obtained by (\ref{19}): \be
\xi_1=\phi^{(1)}\pm \phi^{(2)}\label{eig30} \ .\ee For the other
two eigenvectors the conditions (\ref{conditions}) are not
satisfied and one has to solve the eigenvalue equations by
assuming $\vec n$ given by (\ref{n}) from the very beginning; one
obtains \bea \xi_2&=&\phi^{(1)}\mp\phi^{(2)}\label{eig31}\cr
\xi_3&=&\phi^{(3)}\label{eig32} \ .\eea The eigenvalues for $\vec
n$ along the other 2-fold axes are identical, due to the cubic
symmetry; exactly as in the preceding cases the corresponding
eigenvectors are obtained by (\ref{eig30}), (\ref{eig31}) by
applying the same transformations that change (\ref{n}) to the
unit vectors of the other axes. All these results are summarized
in Table 1 where we have introduced the unit vectors along the
coordinate axes, ${\vec e\,}^{(i)}$.
 We observe again that in the case of propagation along axes of type
$C_2$ there are three different propagation velocities, where for
the other two cases the independent velocities are only two. The
case of propagation along axes of type $C_3$ is of particular
interest since these directions are the same as the directions of
the vectors $\vec q$ specifying the lattice. One of the two
eigenvalues is doubly degenerate, and the non-degenerate
eigenvectors is proportional to \be\sum_{i=1,2,3}
p_i\,\phi^{(i)}\,.\ee The simplest case is for a propagation along
the coordinate axes (type $C_4$), since in this case the matrix
$M$ is diagonal. Also in this case we have a doubly degenerate
eigenvalue with eigenvectors the phonons corresponding to the
plane perpendicular to the direction of propagation, whereas the
non-degenerate eigenvector corresponds to the direction of
propagation. It is interesting to notice that in all cases one of
the eigenvectors is proportional to $\sum_{i} p_i\,\phi^{(i)}$.

\section{Conclusions}
Crystalline superconductivity arises from Cooper pairing between
quarks of non vanishing total momentum, each quark momentum lying
near the respective differing Fermi surfaces. Such a possibility
(LOFF pairing) is known from study of certain condensed matter
systems and recent theoretical studies have made it as highly
probable within the QCD phase diagram. LOFF pairing leads to
spontaneous  breaking of translational and rotational invariance.
Depending on the particular crystal structure a number of phonons
are the lowest energy manifestation of the breaking.

A comparative study of the different possible crystal structures
for the crystalline color superconducting phase of QCD has
recently been published \cite{bowers}, with the conclusion that a
face centered cubic crystal is  the favorite structure. Assuming
this condensate pattern we have derived the expected form of the
effective low energy Lagrangian and the dispersion law for the the
set of the three phonons. The parameters $a$, $b$ and $c$
appearing in the effective lagrangian can be evaluated following
the methods illustrated in ref. \cite{gatto2}. The results will
be given elsewhere.

\begin{center}
\begin{tabular}{||c|c|c||}
\hline\hline &&\\ Momentum & Eigenvalues & Eigenvectors\\&&\\
\hline\hline
 &&\\ \fbox{$C_4$}~~~~~~~~~~~~~~~~~~~~~~~~~~~~~~~~~~~~
 & $(a+b)\,p^2$& $\phi^{(i)}$ \\
 $\vec p = p\, {\vec e\,}^{(i)} $ & $a\, p^2$ &
 $\phi^{(j)}$, $\phi^{(k)}$ ($i\not=j\not=k$)\\
 &&\\ \hline\hline&&\\ \fbox{$C_3$}~~~~~~~~~~~~~~~~~~~~~~~~~~~~~~~~~~~~
 & $(3a+b-c)\,p^2/3$& $-\phi^{(1)}+x\phi^{(2)}+(1-x)\phi^{(3)}$ \\
 $\vec p = p\,({\vec
e\,}^{(1)}+{\vec e\,}^{(2)}+{\vec e\,}^{(3)}) $ & $(3a+b +2c)\,
p^2/3$ &
 $\phi^{(1)}+\phi^{(2)}+\phi^{(3)}$\\
&&\\
 \hline &&\\
& $(3a+b-c)\,p^2/3$& $\phi^{(1)}+x\phi^{(2)}+(1-x)\phi^{(3)}$ \\
 $\vec p = p\,(-{\vec
e\,}^{(1)}+{\vec e\,}^{(2)}+{\vec e\,}^{(3)}) $ & $(3a+b +2c)\,
p^2/3$
& $-\phi^{(1)}+\phi^{(2)}+\phi^{(3)}$\\&&\\
 \hline &&\\
& $(3a+b-c)\,p^2/3$& $\phi^{(1)}+x\phi^{(2)}-(1-x)\phi^{(3)}$ \\
 $\vec p = p\,({\vec
e\,}^{(1)}-{\vec e\,}^{(2)}+{\vec e\,}^{(3)}) $ & $(3a+b +2c)\,
p^2/3$
& $\phi^{(1)}-\phi^{(2)}+\phi^{(3)}$\\&&\\
 \hline &&\\
& $(3a+b-c)\,p^2/3$& $\phi^{(1)}+x\phi^{(2)}+(1+x)\phi^{(3)}$ \\
 $\vec p = p\,({\vec
e\,}^{(1)}+{\vec e\,}^{(2)}-{\vec e\,}^{(3)}) $ & $(3a+b +2c)\,
p^2/3$
& $\phi^{(1)}+\phi^{(2)}-\phi^{(3)}$\\&&\\
 \hline\hline &&\\ \fbox{$C_2$}~~~~~~~~~~~~~~~~~~~~~~~~~~~~~~~~~~~~
 & $a\,p^2$& $\phi^{(k)}$ \\
 $\vec p = p\,({\vec
e\,}^{(i)}+{\vec e\,}^{(j)}) $ & $(2a+b -c)\, p^2/2$ &
 $-\phi^{(i)}+\phi^{(j)}$\\
 $i<j$&$(2a+b +c)\, p^2/2$ &$\phi^{(i)}+\phi^{(j)}$\\&&\\
 \hline &&\\
&$a\,p^2$& $\phi^{(k)}$ \\
 $\vec p = p\,({\vec
e\,}^{(i)}-{\vec e\,}^{(j)}) $ & $(2a+b -c)\, p^2/2$ &
 $\phi^{(i)}+\phi^{(j)}$\\
 $i<j$&$(2a+b +c)\, p^2/2$ &$-\phi^{(i)}+\phi^{(j)}$\\&&\\
 \hline\hline
\end{tabular}
\end{center}
{\bf Table 1} - {\it Eigenvalues and eigenvectors  for  $\vec p$
along the symmetry axes.}

\end{document}